\documentclass{aa}
\usepackage{epsfig}
\usepackage{astron}

\newcommand*{\figszb}{
  \begin{figure}[ht!]
\epsfig{file=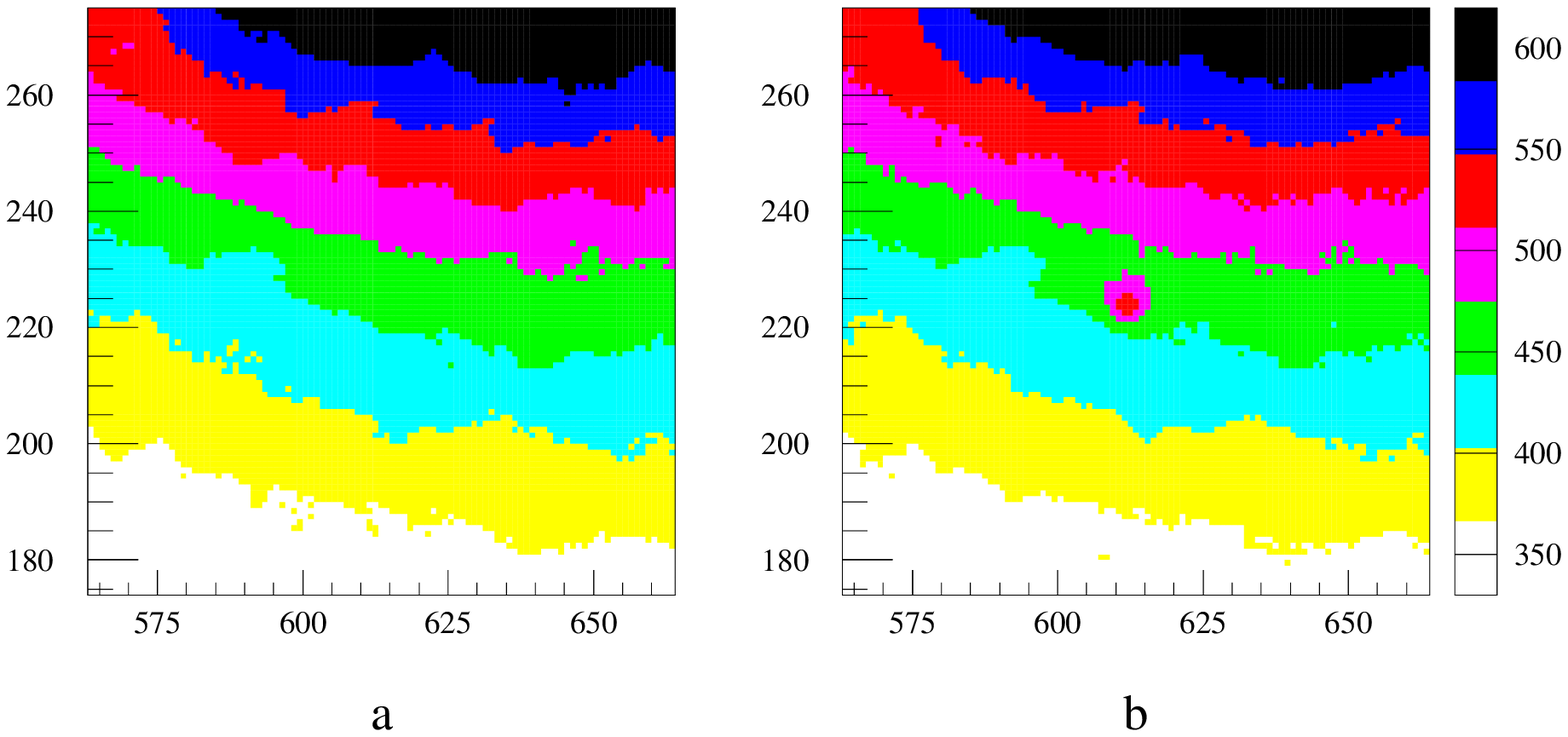,width=.48\textwidth,clip=}
\caption[]{a. Recentred sum of 26x1\,mn exposure B frames onto Z1 position.
a. B 30\,mn exposure image of a 30''x30'' field centred on Z1 at maximum brightness. 
\label{figszb}}  
  \end{figure}}

\newcommand*{\figszblc}{
  \begin{figure}[ht!]
\epsfig{file=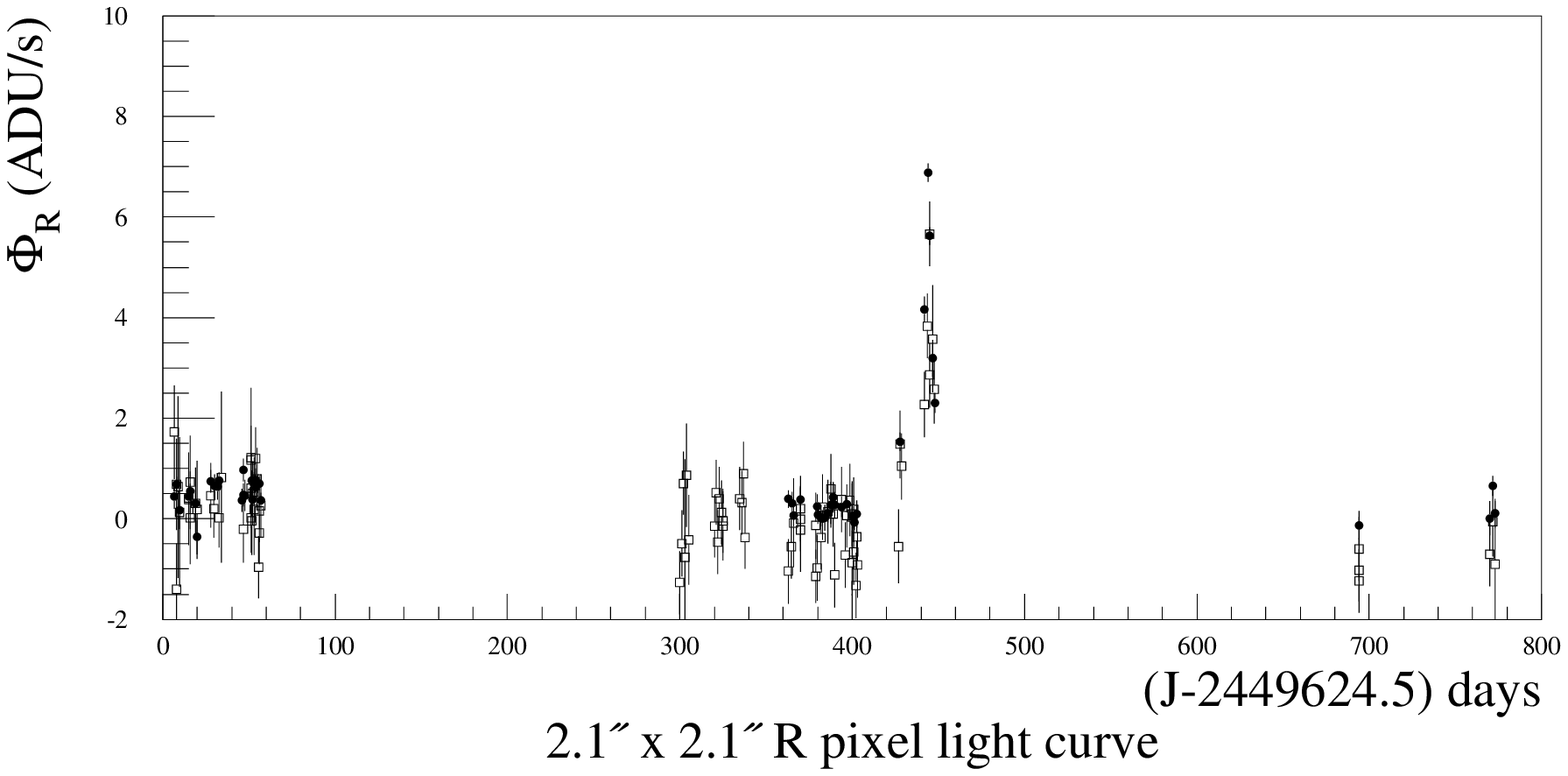,width=.45\textwidth}
\caption[]{
$R$ super-pixel photometry light curve of  AgapeZ1 for all measurements in the central Z (empty dots) and 2$^{nd}$ field (filled dots).
\label{figszblc}}  \end{figure}}

\newcommand*{\figszblcz}{
  \begin{figure}[ht!]
\epsfig{file=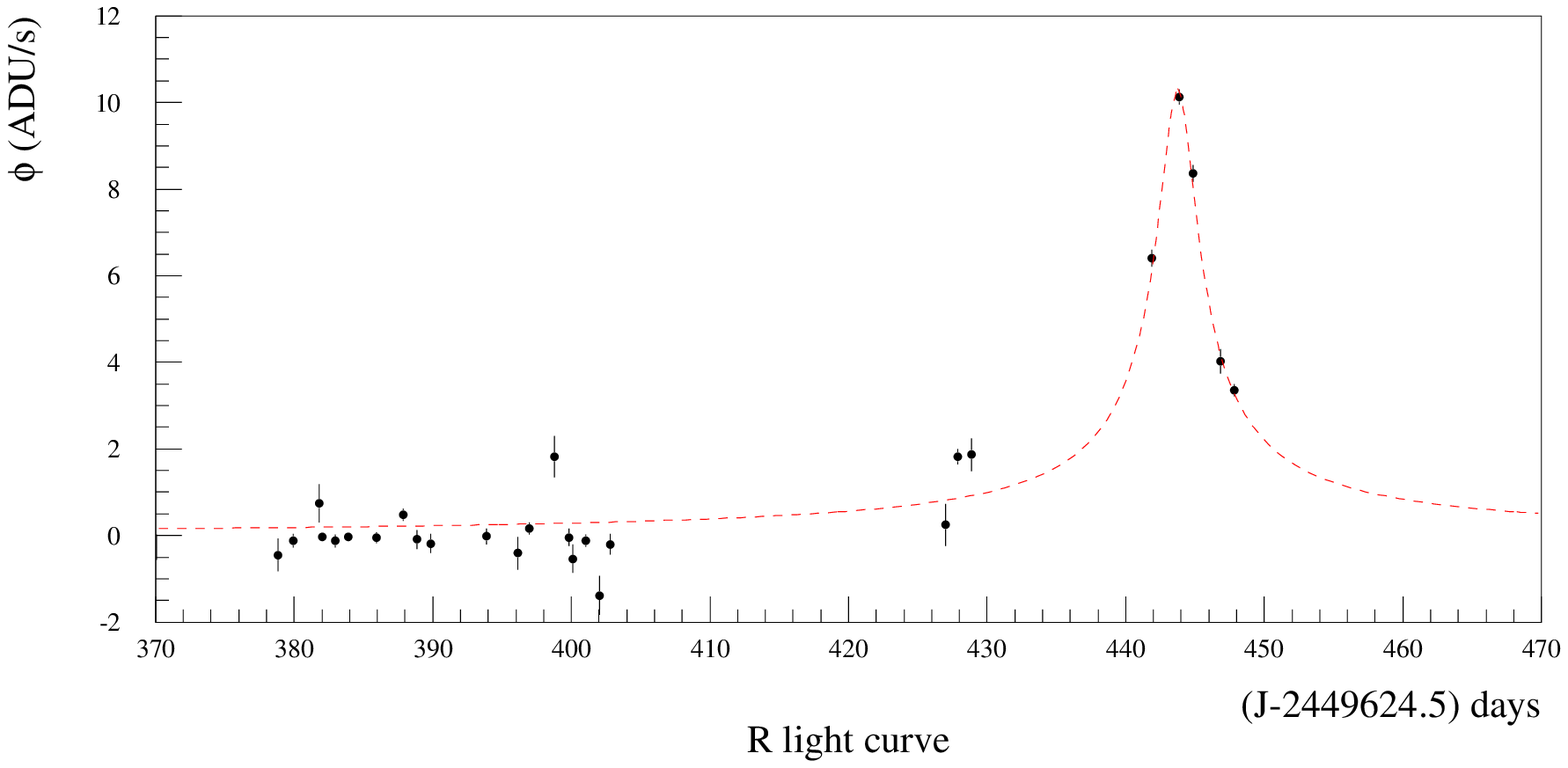,width=.45\textwidth}
\caption[]{
Improved photometry R light curve of Z1 for the event period (1995 season).  The points are
fitted with a degenerate Paczy\'nski curve of parameters $R_{\mathrm max}=18.0$ and $t_{1/2}$=4.8$\pm 0.2$,days.
\label{figszblcz}}  \end{figure}}

\newcommand*{\fighst}{
  \begin{figure}[ht!]
\begin{center}
\psfig{file=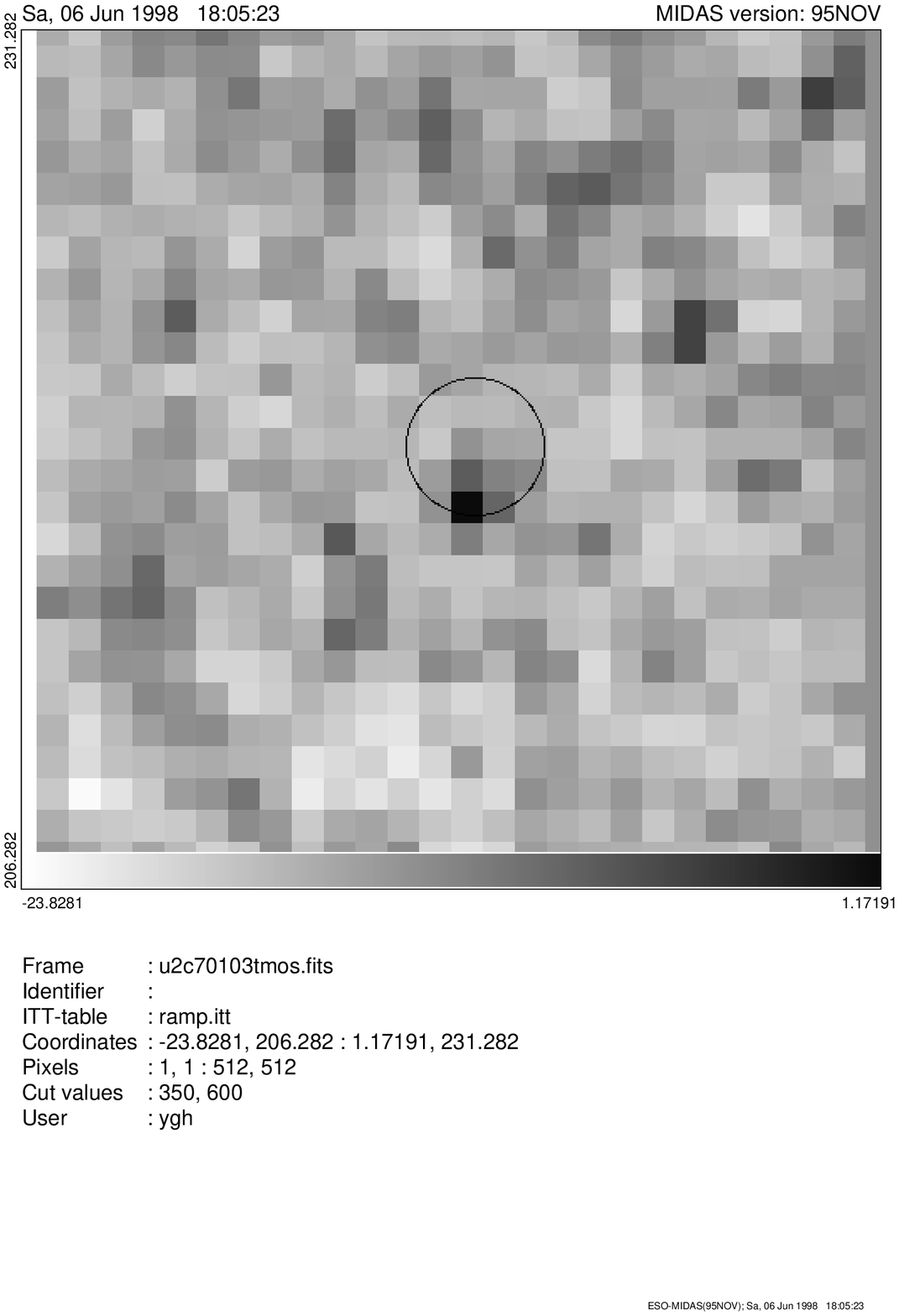,width=.3\textwidth,clip=}
\end{center}
\caption[]{
Negative print of F547M HST image of a 2.6''x2.6'' field centred on HST1 and showing the
3 $\sigma$ error box for Z1 projected position.
\label{fighst}}  \end{figure}}

\newcommand*{\figsevtmc}{
\begin{figure}[ht!]
\psfig{file=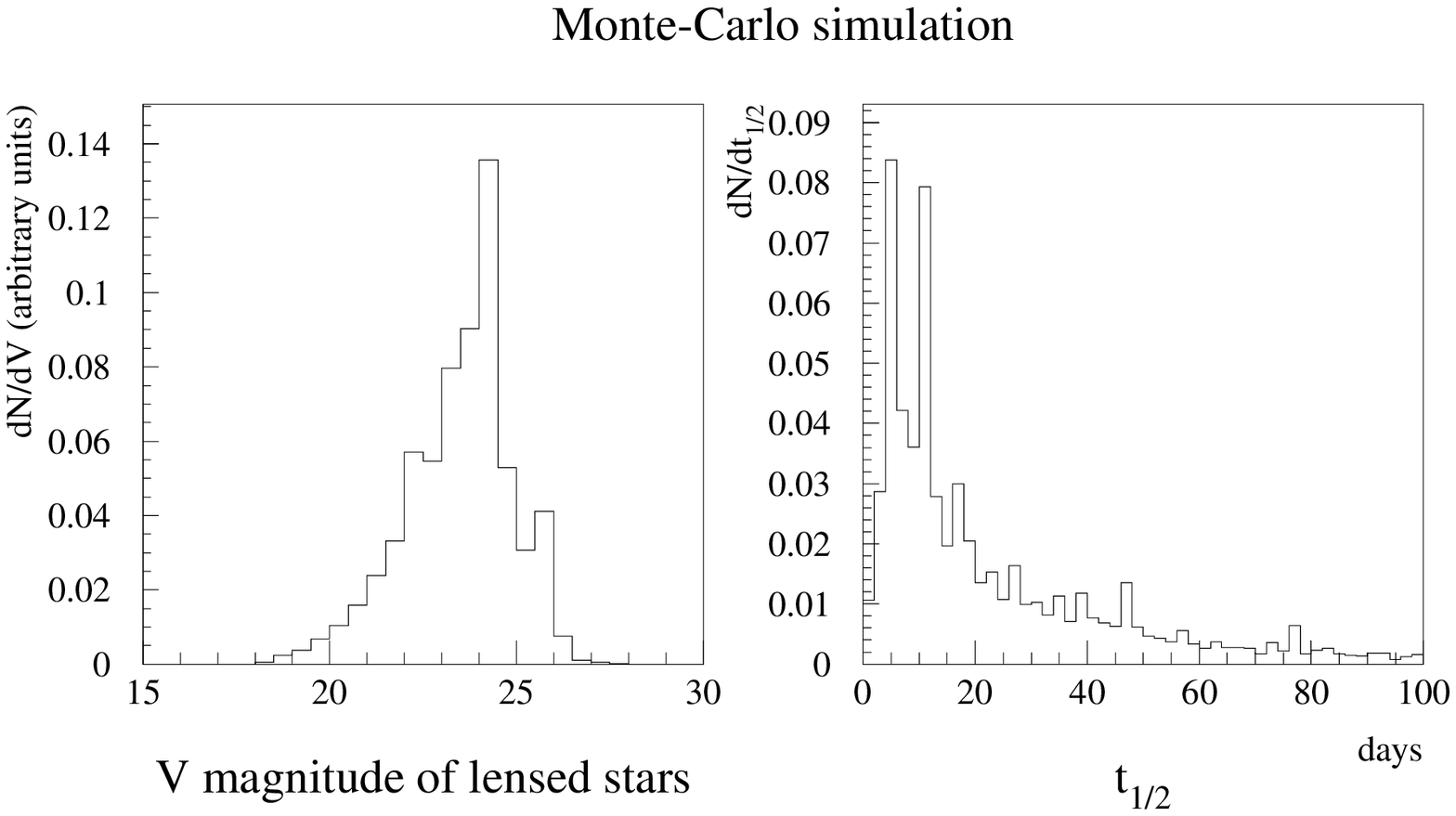,width=.48\textwidth}
\caption[]{
Distributions of the V magnitudes of lensed stars and of the effective
durations of the events expected for  microlensing effects on  M31 stars 
by both  stars of the M31 bulge and Machos belonging to the halo of
M31 or of our own Galaxy. 
\label{figsevtmc}}  \end{figure}}

\newcommand{\beq}{\begin{equation}}
\newcommand{\beqa}{\begin{eqnarray}}
\newcommand{\eeq}{\end{equation}}
\newcommand{\eeqa}{\end{eqnarray}}

\begin{document}
\bibliographystyle{astron}

\thesaurus{06(10.08.01;12.03.3;12.04.1;12.07.1)}

\title{
 AgapeZ1: a large amplification microlensing event or an odd
 variable star towards the inner bulge of  M31
\thanks{Based on data collected with the 2\,m Bernard Lyot Telescope (TBL)
operated by INSU-CNRS and Pic-du-Midi Observatory (USR 5026).
\protect\newline The experiment was funded by IN2P3 and INSU of CNRS}}

\author{ R. Ansari \inst{1} \and M. Auri{\`e}re \inst{2} \and P. Baillon \inst{3}
\and A. Bouquet \inst{4} \and G. Coupinot \inst{2} \and Ch. Coutures \inst{5}
 \and C. Ghesqui{\`e}re \inst{4} \and Y.~Giraud-H{\'e}raud \inst{4} \and
P. Gondolo \inst{6} \and J. Hecquet \inst{7} \and J. Kaplan \inst{4} \and
A. Kim \inst{4} \and Y. Le Du \inst{4} \and A.L. Melchior
\inst{8} \and \\ M. Moniez \inst{1} \and J.P. Picat \inst{7} \and G. Soucail 
\inst{7}}

\titlerunning{AgapeZ1: a large amplification microlensing event or 
  an odd variable star}
\authorrunning{R. Ansari et al.}

\offprints{M. Auri{\`e}re, auriere@obs-mip.fr}

\institute{Laboratoire de l'Acc{\'e}l{\'e}rateur Lin{\'e}aire, Universit{\'e} Paris-Sud,
91405 Orsay, France 
\and Observatoire Midi-Pyr{\'e}n{\'e}es, unit{\'e} associ{\'e}e au CNRS (UMR 5572),
62500 Bagn{\`e}res de Bigorre, France 
\and CERN, 1211 Gen{\`e}ve 23, Switzerland
\and Physique Corpusculaire et Cosmologie, Coll{\`e}ge de France, Laboratoire
associ{\'e} au  CNRS-IN2P3 (UMR 7550), 11 place Marcelin Berthelot, 75231
Paris Cedex 05, France 
\and SPP/DAPNIA, CEN Saclay, 91191 Gif-sur-Yvette, France
\and Max-Planck-Institut f{\"u}r Physik (Werner-Heisenberg-Institut), F{\"o}hringer
Ring 6, 80805 M{\"u}nchen, Germany
\and Observatoire Midi-Pyr{\'e}n{\'e}es, unit{\'e} associ{\'e}e au CNRS (UMR 5572), 14
avenue Belin, 31400 Toulouse, France
\and Astronomy Unit, Queen Mary and Westfield College, Mile End Road, London
E1 4NS, UK}

\date{submitted}

\maketitle

\begin{abstract}
AgapeZ1 is the brightest and the shortest
duration microlensing candidate event found in the
Agape experiment. It occurred
only   $42''$ from the center of M31 at $RA=0^h 42^m 41.47^s$ and
$Dec=41^\circ 16' 39.1''$ (J2000).  Our photometry
shows that the half intensity duration of the event is 4.8\,days and at
maximum brightness we measure a stellar magnitude of $R = 18.0$  ($M_R \sim  -6$) with
$B-R=0.80$ mag color. A search on HST archives produced a single 
resolved star within the projected event position error box. Its
magnitude is $R=22$, and its color is compatible with that of the event
at the $2  \sigma$  level.

If the identification with the HST star is real,
it implies for the event  an amplification of about 4 magnitudes or 40
in brightness. This would lead to an Einstein crossing time radius of about
55 days. AgapeZ1 could be
a bulge/bulge microlensing event involving a binary star. 

The photometric properties of the object exclude
classical M31 variable stars such as miras, novae, dwarf-novae, and bumpers.
However, we cannot rule out the possibility that AgapeZ1
is in fact an odd variable star.

\end{abstract}

\section{Introduction} 
The Agape experiment \cite{agape1} is devoted to the search of dark
matter towards M31. It looks for gravitational
microlensing effects on unresolved
stars by the so-called pixel method \cite{BBGK1,BBGK2}. In the
active field of MACHO microlensing searches
\cite{Paczynski96} only two groups explore the
promising M31 direction \cite{crottsm31,BBGK1}, and though several events with
light curves compatible with microlensing have been presented
\cite{crotts96,YLD98}, all have lacked the strong supporting evidence needed
for them to be classified as true
microlensing events.
The present work describes the properties of AgapeZ1 (hereafter
called Z1),
our brightest and shortest duration candidate which occurred only $42''$ from the
center of M31 in
our central so-called ``Z'' field. 
The observational interest of our central field is manifold. It was
observed at least once each of the 79 observing
nights in R, and 30 nights in B.
The central region of M31 has been
observed by HST allowing for high-resolution archival
searches for the quiescent sources of detected events. The central
region of M31 is also of great astrophysical interest since it contains a huge
number of stars. It is thus in this direction that the
greatest number of microlensing events is expected to occur. 
However, because of the high star background level, only
those with the highest amplification parameters will be
resolved. A large number of variable stars, including exotic objects,  may also be expected.

\section{TBL observations and photometry \label{obsphot}}
The Agape observations were made at the 2\,m Bernard Lyot telescope (TBL) of
the Pic du Midi Observatory with the F/8 spectro-reducer ISARD. A
thin Tektronik 1024x1024 CCD was used with a useful field of $4' \times 4.5'$
with $0.3 ''$ pixels. The exposure times were generally 
1\,min in both the B and R passbands for the (Z) central field and
30(20)\,min in the B(R) passband for the 6 other fields investigated
by the experiment  \cite{agape1}.
We have 93 (30) R (B)  exposures for field ``Z'' and 70 (33) R (B) exposures  on the edge of a second field.
The observing campaign ran from 1994 to 1996.

The Agape detection procedure is described in Ansari et al. \cite*{agape1}.
It is based on  the photometry in super-pixel (grouping of 7x7 pixels)
with sides roughly two times the 
standard seeing. These super-pixels 
are photometrically normalised to a reference
frame and corrected for seeing variations.
The light curves for each
super-pixel are analysed, yielding
over 2000 variable objects.
Of these, the 61 with only a single bump are then fitted with
degenerate Paczy\'nski curves \cite{wozniak97}.
Selecting those with $\chi^2/dof < 1.5$ leaves only 19 light-curves. After a cut on the color and the event duration (simulations described in section 4.1 show that 70\,\% of expected microlensing events should have half intensity duration shorter than 40 days) we are left with only two candidates.\figszb
Z1 is brighter, the shorter time-scale event and is
located in the central bulge (Fig. \ref{figszb}). In our non central field containing Z1, it lies close to the edge
where shadowing occurs which
causes some systematic
photometric uncertainty. However, because of their larger exposure
times, the precision
of the second field observations is greater than for the Z field
ones. Fig. \ref{figszblc}  shows the super-pixel light
curve found by Agape for the object Z1.  The days correspond to
$J-2449624.5$ where J is the julian date.

To study the selected candidates, we developed a sophisticated
photometry which will be described in a forthcoming paper \cite{PB98}.
Our procedure belongs to the so-called image subtraction technics, already  used by
Tomaney and Crotts \cite*{tomaney96}  and Alard and Lupton \cite*{alard98}. It is based
on a global fit of one PSF (10x10 parameters) for each image and a
unique  reference background field (200x200 parameters). As it  takes seeing effects more efficiently into account than
does the super-pixel photometry described in Ansari et al. \cite*{agape1} we found its results more accurate, with however still $15\%$ systematic uncertainty . 

 \figszblc \figszblcz

Fig. \ref{figszblcz} shows
an enlargement of the improved photometry light curve of Z1 at the time of the
event, averaging the star fluxes measured on the 2 fields when
available. Apart from a (significant) bump at day $\approx 428$, a good fit
of a Paczy\'nski curve ($t_{1/2}$=4.8\,days) is obtained (data for 1994 and 1996 seasons  are also used for the fit). At maximum, on 16 december 1995,
the R magnitude is found to be $R$=18.0  and the color
$B-R=0.80$. We have four color measurements during the event. The color at maximum has accurate precision (0.05 mag. stat.) since the corresponding B image is a 30 min exposure. Because of the faintness of the star or short exposures, the three
other measurements cannot be used to tightly constrain achromaticity.

\section{HST observations of the Z1 field}

The Z1 event positional
error box lies on a series of HST WFPC2 archive taken on 9 September 1994 as part of a single
observing program\footnote{Based on observations made with the NASA/ESA Hubble Space Telescope,
  obtained from the data archive at the Space Telescope Science
  Institute. STScI is operated by the Association of Universities for
  Research in Astronomy, Inc. under the NASA contract NAS 5-26555.}
 (The PI was R. Bohlin from STScI). There were no change in  pointing between each
exposure. We studied one 2300\,s image taken with the F656N filter and
two co-added images 
with an effective exposure of 1200\,s taken with the F547M filter.

We have computed the spatial transformation between the Agape and HST fields
with a least square reduction based on 10 common stars. The standard
deviation for the projection accuracy of the standard stars is
{$0.06''$} and is  mainly due to the uncertainties on the position of the
stars in the Agape fields.
We have projected the position of Z1 onto the HST images with a
3\,$\sigma$ uncertainty of $0.18''$ and found a
faint star ($R \sim 22$  mag)  {$0.14''$} away from the projected position. It is the only
resolved star on the HST image nearer than $0.4''$ from the Z1
projection. We
call this star HST1. Fig.~\ref{fighst} shows a negative print of the
HST field for the Z1 projected  region.\fighst

We use the DAOPHOT
package \cite{stetson87} to perform PSF fitting on the star. Since the
field is crowded and there are no bright stars near our candidate, we
used the theoretical PSF of  TINYTIM.
The standard magnitudes from the PSF photometry are $m_{F656N}=21.6\pm
0.2$ and $m_{F547M}=21.9\pm 0.1$ which are consistent  within error from the results obtained from aperture photometry.

\section{Interpretation \label{interpretation}}
\subsection{ AgapeZ1 as a microlensing event}

\figsevtmc
On Fig. \ref{figsevtmc} we show two plots obtained from Monte-Carlo simulations which include
the known characteristics of the two galaxies (M31 and Milky Way), and, for each,
 an isothermal halo filled with 0.5\,$M_\odot$
machos. With respect to simulations described in
\cite{agape1}, 0.6\,$M_\odot$ M31 bulge lenses are added. Simulations give the distributions of  the V magnitude of lensed stars and of the effective
duration ($t_{1/2}$ defined as the FWHM of the amplification peak in the lightcurve) of the event expected for a microlensing effect detected with 
the same criteria as in our selection process. 

If Z1 is interpreted as
being
 due to a microlensing amplification of HST1, the event
characteristics ($V \sim 22$ and $t_{1/2}=4.8$\,days)
are fully compatible with those expected considering 
these simulations. In this case the magnification is of 4 magnitudes or 40 in brightness and
the Einstein
radius crossing time is about 55 days . This is typical for a
microlensing event between bulge-bulge stars with a mass of
0.6\,$M_\odot$ which is expected whatever is the nature of the halo or for a microlensing event between a halo macho of 0.5\,$M_\odot$ and a M31 star.

Now, in this hypothesis, HST1 and Z1 must have
the same color and  the same spectral type. After correcting for the galactic extinction of  $E_{B-V}=0.08$ \cite{vandenbergh91} with the extinction model of
Cardelli, Clayton \& Mathis \cite*{cardelli89}, we find that Z1 has the $B-R$ color of an
F5 star \cite{allen73}. The color and magnitude of HST1 are consistent also with an F5II star at the $2  \sigma$  level.
However F5II stars are rare and only some tens are found in huge
spectroscopic catalogs \cite{houk78}. They
correspond to a very short stage in stellar evolution of massive
stars. For example, using a  ``Geneve"  model \cite{schaller92} we
find that it would correspond to a subgiant of 4 $M_\odot$ ( between the main
sequence and the helium flash). 
 
The color and magnitude of HST1 can also be attributed to a highly reddened supergiant. However, such a large extinction is  unlikely in M31 considering Han \cite*{han96} measurement of a uniform  $A_{V}=0.24$ disk extinction. We thus prefer the identification of HST1 as an F5II star.

The fit with a Paczy\'nski model ($t_{1/2}$=4.8\,days) is good for all
points except for a statistically significant bump two days before the rapid
rise to maximum. The shape of this lightcurve could be
explained by the presence of a binary source \cite{griest92} or a
  binary lens \cite{distefano97}. The binary source hypothesis could explain 
  the odd color of Z1/HST1 and the possible difference of color between the two objects.
 
Finally, Z1 could be a microlensing amplification of a fainter star, blended or not with HST1. In this case the source would be at least 1 magnitude  fainter than HST1, and the amplification greater than 100.

\subsection{ AgapeZ1 as a variable star:} 

M31 variable stars could mimic
microlensing events.
For example, Crotts and Tomaney \cite*{crotts96} point out the
possible pollution of their sample of candidates by very long period
Miras, some of  which they have already discarded. In the case of Z1, its
blue color definitively excludes this hypothesis.

Della Valle and Livio (1996) \nocite{dellavalle96}
explored the possibility that dwarf
novae could contaminate microlensing survey samples. For observed dwarf novae,
colour  ranges between $B-V = -0.1$ and $B-V = +0.6$  and main outburst amplitude ranges between 2 and 5 magnitudes  \cite{warner95}. Thus the
colors of HST1 and Z1 as well as the amplitude of the event are
consistent with a dwarf nova outburst. However, the
quiescent absolute magnitude of dwarf novae is around 7 with a rather
large range \cite{warner95}. If HST1 and Z1 correspond to a
dwarf nova, the object would be in the foreground, well outside
M31 and in the Galactic halo within  10\,kpc   from
the sun . The existence of such an object
 exactly projected towards the inner bulge of M31 is rather unlikely. There
exists a broad relation between outburst amplitude and outburst interval for
dwarf novae \cite{warner95}: for a 4 magnitude amplitude, one can expect
outbursts to occur  with intervals smaller than 100 days. In this case, the repetition of the Z1 event could be observable with follow-up observations.

Bumpers are variable stars which were detected  by the MACHO
experiment \cite{alcock96}. These objects have small amplitudes,  unlike Z1
event.

Although the overall appearance of Z1 is similar to that of  a nova, its faint
magnitude would imply a long  rate of decline while we observe a rapid
one (0.25 magnitude per day observed, for 0.02-0.04 magnitude per day expected from the relation
established by Capaccioli et al. \cite*{capaccioli89} for M31 novae). Reconciliation
with the Capaccioli et al. trend would require Z1 to have a reddening of around
2 magnitudes in the visible which would imply a $E_{B-R}$ reddening of
about 1 magnitude. The Agape experiment
observed ten M31 novae, two being in the   ``Z field", and nine having 
$B-R$ colors.  All the novae for which the respective relevant data are
available follow the Capaccioli et al. \cite*{capaccioli89} trend and/or they
have a $B-R$ color near
maximum in the range 0.4-0.6 (apart from one which is strongly  reddened
near maximum). The
color of $B-R=0.80$ at maximum for Z1 is thus not what would be  expected for an M31
nova reddened by 2 magnitudes in the visible.

\section{Conclusion \label{conclusion}}
Our work shows that the AgapeZ1 event could be due to the
gravitational amplification of a F5II color binary object
corresponding to HST1 with an Einstein radius crossing time of 55
days.  On the other hand the photometric properties of Z1 are incompatible
with those of a classical M31 variable star. The foreground dwarf
novae hypothesis appears unlikely. However, the inner bulge of M31 may
be the site for rather odd objects. We have thus compared the position of Z1 with those of already known
exotic objects including the 1885 supernovae  \cite{devaucouleurs85},
the X-ray sources \cite{primini93,trinchieri91}, and novae observed up
to the inner bulge by \cite{ciardullo87}.    
Z1 is
located  about 6'' from the position of  the transient X-ray source
E47 but the chance of association is weak. However, Z1 could be  an unknown kind of cataclysmic variable. New Z field  Agape observations are on the way to monitor for a recurrence of Z1 and to search  for  similar
objects in the bulge of  M31.

\bibliography{mnemoaa,lensing,newagp}

\end{document}